\documentclass[preprint2,bibnotes,secnumarabic]{aastex}

\usepackage{graphicx}
\usepackage{multirow}
\usepackage{natbib}
\usepackage{float}

\shorttitle{Low Multiplicity Burst Search} \shortauthors{Aharmim \etal}

\begin{document}

\title{Low Multiplicity Burst Search at the Sudbury Neutrino Observatory}

\newcommand{\alta}{Department of Physics, University of 
Alberta, Edmonton, Alberta, T6G 2R3, Canada}
\newcommand{\ubc}{Department of Physics and Astronomy, University of 
British Columbia, Vancouver, BC V6T 1Z1, Canada}
\newcommand{\bnl}{Chemistry Department, Brookhaven National 
Laboratory,  Upton, NY 11973-5000}
\newcommand{\carleton}{Ottawa-Carleton Institute for Physics, Department of Physics, Carleton University, Ottawa, Ontario K1S 5B6, Canada}
\newcommand{\uog}{Physics Department, University of Guelph,  
Guelph, Ontario N1G 2W1, Canada}
\newcommand{\lu}{Department of Physics and Astronomy, Laurentian 
University, Sudbury, Ontario P3E 2C6, Canada}
\newcommand{\lbnl}{Institute for Nuclear and Particle Astrophysics and 
Nuclear Science Division, Lawrence Berkeley National Laboratory, Berkeley, CA 94720}
\newcommand{\lanl}{Los Alamos National Laboratory, Los Alamos, NM 87545}
\newcommand{\lanla}{Present address: Los Alamos National Laboratory, Los Alamos, NM 87545}
\newcommand{\oxford}{Department of Physics, University of Oxford, 
Denys Wilkinson Building, Keble Road, Oxford OX1 3RH, UK}
\newcommand{\penn}{Department of Physics and Astronomy, University of 
Pennsylvania, Philadelphia, PA 19104-6396}
\newcommand{\queens}{Department of Physics, Queen's University, 
Kingston, Ontario K7L 3N6, Canada}
\newcommand{\uw}{Center for Experimental Nuclear Physics and Astrophysics, 
and Department of Physics, University of Washington, Seattle, WA 98195}
\newcommand{\uta}{Department of Physics, University of Texas at Austin, Austin, TX 78712-0264}
\newcommand{\triumf}{TRIUMF, 4004 Wesbrook Mall, Vancouver, BC V6T 2A3, Canada}
\newcommand{\ralimp}{Rutherford Appleton Laboratory, Chilton, Didcot OX11 0QX, UK}
\newcommand{\uci}{Department of Physics, University of California, Irvine, CA 92717}
\newcommand{\snoi}{SNOLAB, Lively, ON P3Y 1N2, Canada}
\newcommand{\lbla}{Present address: Lawrence Berkeley National Laboratory, Berkeley, CA}
\newcommand{\llnl}{Present address: Lawrence Livermore National Laboratory, Livermore, CA}
\newcommand{\iusb}{Present address: Department of Physics and Astronomy, Indiana University, South Bend, IN}
\newcommand{\fnal}{Present address: Fermilab, Batavia, IL}
\newcommand{\uo}{Present address: Department of Physics and Astronomy, University of Oregon, Eugene, OR}
\newcommand{\hu}{Present address: Department of Physics, Hiroshima University, Hiroshima, Japan}
\newcommand{\slac}{Present address: Stanford Linear Accelerator Center, Menlo Park, CA}
\newcommand{\mac}{Present address: Department of Physics, McMaster University, Hamilton, ON}
\newcommand{\doe}{Present address: US Department of Energy, Germantown, MD}
\newcommand{\lund}{Present address: Department of Physics, Lund University, Lund, Sweden}
\newcommand{\mpi}{Present address: Max-Planck-Institut for Nuclear Physics, Heidelberg, Germany}
\newcommand{\uom}{Present address: Ren\'{e} J.A. L\'{e}vesque Laboratory, Universit\'{e} de Montr\'{e}al, Montreal, PQ}
\newcommand{\cwru}{Present address: Department of Physics, Case Western Reserve University, Cleveland, OH}
\newcommand{\pnnl}{Present address: Pacific Northwest National Laboratory, Richland, WA}
\newcommand{\uc}{Present address: Department of Physics, University of Chicago, Chicago, IL}
\newcommand{\mitt}{Laboratory for Nuclear Science, Massachusetts Institute of Technology, Cambridge, MA 02139} 
\newcommand{\ucsd}{Present address: Department of Physics, University of California at San Diego, La Jolla, CA }
\newcommand{	\lsu	}{Department of Physics and Astronomy, Louisiana State University, Baton Rouge, LA 70803}
\newcommand{\imp}{Imperial College, London, UK} %SW7 2AZ
\newcommand{\ucia}{Department of Physics, University of California, Irvine, CA}
\newcommand{\suss}{Present address: Department of Physics and Astronomy, University of Sussex, Brighton, UK} %BN1 9QH
\newcommand{	\lifep	}{Laborat\'{o}rio de Instrumenta\c{c}\~{a}o e F\'{\i}sica Experimental de
Part\'{\i}culas, Av. Elias Garcia 14, 1$^{\circ}$,  Lisboa, Portugal} %1000-149
\newcommand{\hku}{Department of Physics, The University of Hong Kong, Hong Kong.}
\newcommand{\aecl}{Atomic Energy of Canada, Limited, Chalk River Laboratories, Chalk River, ON , Canada} %K0J 1J0
\newcommand{\nrc}{National Research Council of Canada, Ottawa, ON, Canada} %K1A 0R6
\newcommand{\princeton}{Department of Physics, Princeton University, Princeton, NJ} %08544
\newcommand{\birkbeck}{Birkbeck College, University of London, Malet Road, London, UK} %WC1E 7HX
\newcommand{\snoia}{SNOLAB, Sudbury, ON, Canada} %P3Y 1M3
\newcommand{\uba}{Present address: University of Buenos Aires, Argentina}
\newcommand{\hvd}{Present address: Department of Physics, Harvard University, Cambridge, MA}
\newcommand{\pny}{Present address: Goldman Sachs, 85 Broad Street, New York, NY}
\newcommand{\pnv}{Present address: Remote Sensing Lab, PO Box 98521, Las Vegas, NV} %89193
\newcommand{\psis}{Present address: Paul Schiffer Institute, Villigen, Switzerland}
\newcommand{\liverpool}{Present address: Department of Physics, University of Liverpool, Liverpool, UK}
\newcommand{\uto}{Present address: Department of Physics, University of Toronto, Toronto, ON, Canada}
\newcommand{\uwisc}{Present address: Department of Physics, University of Wisconsin, Madison, WI}
\newcommand{\psu}{Present address: Department of Physics, Pennsylvania State University,
     University Park, PA}
\newcommand{\anl}{Present address: Department of Mathematics and Computer Science, Argonne
     National Laboratory, Lemont, IL}
\newcommand{\cornell}{Present address: Department of Physics, Cornell University, Ithaca, NY}
\newcommand{\tufts}{Present address: Department of Physics and Astronomy, Tufts University, Medford, MA}
\newcommand{\ucd}{Present address: Department of Physics, University of California, Davis, CA}
\newcommand{\unc}{Present address: Department of Physics, University of North Carolina, Chapel Hill, NC}
\newcommand{\dresden}{Present address: Institut f\"{u}r Kern- und Teilchenphysik, Technische Universit\"{a}t Dresden,   Dresden, Germany} %01069
\newcommand{\isu}{Present address: Department of Physics, Idaho State University, Pocatello, ID}
\newcommand{\qmul}{Present address: Dept. of Physics, Queen Mary University, London, UK}
\newcommand{\ucsb}{Present address: Dept. of Physics, University of California, Santa Barbara, CA}
\newcommand{\cern}{Present address: CERN, Geneva, Switzerland}
\newcommand{\casa}{Present address: Center for Astrophysics and Space Astronomy, University
of Colorado, Boulder, CO}  %80309
\newcommand{\aasu}{Present address: Department of Chemistry and Physics, Armstrong Atlantic State University, Savannah, GA}
\newcommand{\susel}{Present address: Sanford Laboratory at Homestake, Lead, SD}  %57754
\newcommand{\uwa}{Present address: Center for Experimental Nuclear Physics and Astrophysics, 
and Department of Physics, University of Washington, Seattle, WA} %98195
\newcommand{\queensa}{Present address: Department of Physics, Queen's University, 
Kingston, Ontario, Canada} % K7L 3N6
\newcommand{\ntu}{Present address: Center of Cosmology and Particle Astrophysics, National Taiwan University, Taiwan}
\newcommand{\berlin}{Present address: Institute for Space Sciences, Freie Universit\"{a}t Berlin,
Leibniz-Institute of Freshwater Ecology and Inland Fisheries, Germany}
\newcommand{\bhsu}{Present address: Black Hills State University, Spearfish, SD} %57799-9003

\author{
B.~Aharmim\altaffilmark{\ref{lu}},
S.\,N.~Ahmed\altaffilmark{\ref{queens}},
A.\,E.~Anthony\altaffilmark{\ref{uta},\ref{casa}},
N.~Barros\altaffilmark{\ref{lifep}},
E.\,W.~Beier\altaffilmark{\ref{penn}},
A.~Bellerive\altaffilmark{\ref{carleton}},
B.~Beltran\altaffilmark{\ref{alta}},
M.~Bergevin\altaffilmark{\ref{lbnl},\ref{uog}},
S.\,D.~Biller\altaffilmark{\ref{oxford}},
K.~Boudjemline\altaffilmark{\ref{carleton},\ref{queens}},
M.\,G.~Boulay\altaffilmark{\ref{queens}},
B.~Cai\altaffilmark{\ref{queens}},
Y.\,D.~Chan\altaffilmark{\ref{lbnl}},
D.~Chauhan\altaffilmark{\ref{lu}},
M.~Chen\altaffilmark{\ref{queens}},
B.\,T.~Cleveland\altaffilmark{\ref{oxford}},
X.~Dai\altaffilmark{\ref{queens},\ref{oxford},\ref{carleton}},
H.~Deng\altaffilmark{\ref{penn}},
J.~Detwiler\altaffilmark{\ref{lbnl}},
G.~Doucas\altaffilmark{\ref{oxford}},
P.-L.~Drouin\altaffilmark{\ref{carleton}},
F.\,A.~Duncan\altaffilmark{\ref{snoi},\ref{queens}},
M.~Dunford\altaffilmark{\ref{penn},\ref{cern}},
E.\,D.~Earle\altaffilmark{\ref{queens}},
S.\,R.~Elliott\altaffilmark{\ref{lanl},\ref{uw}},
H.\,C.~Evans\altaffilmark{\ref{queens}},
G.\,T.~Ewan\altaffilmark{\ref{queens}},
J.~Farine\altaffilmark{\ref{lu},\ref{carleton}},
H.~Fergani\altaffilmark{\ref{oxford}},
F.~Fleurot\altaffilmark{\ref{lu}},
R.\,J.~Ford\altaffilmark{\ref{snoi},\ref{queens}},
J.\,A.~Formaggio\altaffilmark{\ref{mitt},\ref{uw}},
N.~Gagnon\altaffilmark{\ref{uw},\ref{lanl},\ref{lbnl},\ref{oxford}},
J.\,TM.~Goon\altaffilmark{\ref{lsu}},
K.~Graham\altaffilmark{\ref{carleton},\ref{queens}},
E.~Guillian\altaffilmark{\ref{queens}},
S.~Habib\altaffilmark{\ref{alta}},
R.\,L.~Hahn\altaffilmark{\ref{bnl}},
A.\,L.~Hallin\altaffilmark{\ref{alta}},
E.\,D.~Hallman\altaffilmark{\ref{lu}},
P.\,J.~Harvey\altaffilmark{\ref{queens}},
R.~Hazama\altaffilmark{\ref{uw},\ref{hu}},
W.\,J.~Heintzelman\altaffilmark{\ref{penn}},
J.~Heise\altaffilmark{\ref{ubc},\ref{lanl},\ref{queens},\ref{susel}},
R.\,L.~Helmer\altaffilmark{\ref{triumf}},
A.~Hime\altaffilmark{\ref{lanl}},
C.~Howard\altaffilmark{\ref{alta}},
M.~Huang\altaffilmark{\ref{uta},\ref{lu},\ref{ntu}},
B.~Jamieson\altaffilmark{\ref{ubc}},
N.\,A.~Jelley\altaffilmark{\ref{oxford}},
M.~Jerkins\altaffilmark{\ref{uta}},
J.\,R.~Klein\altaffilmark{\ref{uta},\ref{penn}},
M.~Kos\altaffilmark{\ref{queens}},
C.~Kraus\altaffilmark{\ref{queens},\ref{lu}},
C.\,B.~Krauss\altaffilmark{\ref{alta}},
T.~Kutter\altaffilmark{\ref{lsu}},
C.\,C.\,M.~Kyba\altaffilmark{\ref{penn},\ref{berlin}},
J.~Law\altaffilmark{\ref{uog}},
K.\,T.~Lesko\altaffilmark{\ref{lbnl}},
J.\,R.~Leslie\altaffilmark{\ref{queens}},
J.\,C.~Loach\altaffilmark{\ref{oxford},\ref{lbnl}},
R.~MacLellan\altaffilmark{\ref{queens}},
S.~Majerus\altaffilmark{\ref{oxford}},
H.\,B.~Mak\altaffilmark{\ref{queens}},
J.~Maneira\altaffilmark{\ref{lifep}},
R.~Martin\altaffilmark{\ref{queens},\ref{lbnl}},
N.~McCauley\altaffilmark{\ref{penn},\ref{oxford},\ref{liverpool}},
A.\,B.~McDonald\altaffilmark{\ref{queens}},
M.\,L.~Miller\altaffilmark{\ref{mitt},\ref{uw}},
B.~Monreal\altaffilmark{\ref{mitt},\ref{ucsb}},
J.~Monroe\altaffilmark{\ref{mitt}},
B.\,G.~Nickel\altaffilmark{\ref{uog}},
A.\,J.~Noble\altaffilmark{\ref{queens},\ref{carleton}},
H.\,M.~O'Keeffe\altaffilmark{\ref{oxford},\ref{queensa}},
N.\,S.~Oblath\altaffilmark{\ref{uw},\ref{mitt}},
G.\,D.~Orebi Gann\altaffilmark{\ref{oxford},\ref{penn}},
S.\,M.~Oser\altaffilmark{\ref{ubc}},
R.\,A.~Ott\altaffilmark{\ref{mitt}},
S.\,J.\,M.~Peeters\altaffilmark{\ref{oxford},\ref{suss}},
A.\,W.\,P.~Poon\altaffilmark{\ref{lbnl}},
G.~Prior\altaffilmark{\ref{lbnl},\ref{cern}},
S.\,D.~Reitzner\altaffilmark{\ref{uog}},
K.~Rielage\altaffilmark{\ref{lanl},\ref{uw}},
B.\,C.~Robertson\altaffilmark{\ref{queens}},
R.\,G.\,H.~Robertson\altaffilmark{\ref{uw}},
M.\,H.~Schwendener\altaffilmark{\ref{lu}},
J.\,A.~Secrest\altaffilmark{\ref{penn},\ref{aasu}},
S.\,R.~Seibert\altaffilmark{\ref{uta},\ref{lanl},\ref{penn}},
O.~Simard\altaffilmark{\ref{carleton}},
J.\,J.~Simpson\altaffilmark{\ref{uog}},
D.~Sinclair\altaffilmark{\ref{carleton},\ref{triumf}},
P.~Skensved\altaffilmark{\ref{queens}},
T.\,J.~Sonley\altaffilmark{\ref{mitt},\ref{queensa}},
L.\,C.~Stonehill\altaffilmark{\ref{lanl},\ref{uw}},
G.~Te\v{s}i\'{c}\altaffilmark{\ref{carleton}},
N.~Tolich\altaffilmark{\ref{uw}},
T.~Tsui\altaffilmark{\ref{ubc}},
R.~Van~Berg\altaffilmark{\ref{penn}},
B.\,A.~VanDevender\altaffilmark{\ref{uw},\ref{pnnl}},
C.\,J.~Virtue\altaffilmark{\ref{lu}},
H.~Wan~Chan~Tseung\altaffilmark{\ref{oxford},\ref{uw}},
P.\,J.\,S.~Watson\altaffilmark{\ref{carleton}},
N.~West\altaffilmark{\ref{oxford}},
J.\,F.~Wilkerson\altaffilmark{\ref{uw},\ref{unc}},
J.\,R.~Wilson\altaffilmark{\ref{oxford},\ref{qmul}},
A.~Wright\altaffilmark{\ref{queens}},
M.~Yeh\altaffilmark{\ref{bnl}},
F.~Zhang\altaffilmark{\ref{carleton}},
K.~Zuber\altaffilmark{\ref{oxford},\ref{dresden}}							}
\author{SNO Collaboration}

\altaffiltext{1}{  \label{alta} \alta   }
\altaffiltext{2}{  \label{ubc} \ubc  }
\altaffiltext{3}{  \label{bnl} \bnl  } 
\altaffiltext{4}{  \label{carleton} \carleton } 
\altaffiltext{5}{  \label{uog} \uog   } 
\altaffiltext{6}{  \label{lu} \lu  }
\altaffiltext{7}{  \label{lbnl} \lbnl  } 
\altaffiltext{8}{  \label{lanl} \lanl   } 
\altaffiltext{9}{  \label{lifep} \lifep  } 
\altaffiltext{10}{  \label{lsu} \lsu   } 
\altaffiltext{11}{  \label{mitt} \mitt   } 
\altaffiltext{12}{  \label{oxford} \oxford  } 
\altaffiltext{13}{  \label{penn} \penn }
\altaffiltext{14}{  \label{queens} \queens   }
\altaffiltext{15}{  \label{ralimp} \ralimp}
\altaffiltext{16}{  \label{uw} \uw   } 
\altaffiltext{17}{  \label{snoi} \snoi  } 
\altaffiltext{18}{  \label{uta} \uta}
\altaffiltext{19}{  \label{triumf} \triumf  } 

\altaffiltext{20}{ \label{ntu} \ntu}
\altaffiltext{21}{  \label{casa} \casa}
\altaffiltext{22}{ \label{uc} \uc}
\altaffiltext{23}{ \label{hu} \hu}
\altaffiltext{24}{ \label{susel} \susel}
\altaffiltext{25}{ \label{liverpool} \liverpool}
\altaffiltext{26}{ \label{ucsb} \ucsb}
\altaffiltext{27}{ \label{queensa} \queensa}
\altaffiltext{28}{ \label{suss} \suss}
\altaffiltext{29}{ \label{cern} \cern}
\altaffiltext{30}{ \label{aasu} \aasu}
\altaffiltext{31}{ \label{unc} \unc}
 %\altaffiltext{32}{  \label{imp} Alternate address: Imperial College, London SW7 2AZ, UK }  
\altaffiltext{32}{ \label{qmul} \qmul}
\altaffiltext{33}{ \label{dresden} \dresden}
\altaffiltext{34}{ \label{pnnl} \pnnl}
\altaffiltext{35}{ \label{berlin} \berlin}
%\altaffiltext{34}{ \label{bhsu} \bhsu}

\shorttitle{SNO Burst} \shortauthors{Aharmim \etal}

\begin{abstract}
Results are reported from a search for low-multiplicity neutrino
bursts in the Sudbury Neutrino Observatory (SNO).  Such bursts could
indicate detection of a nearby core-collapse supernova explosion.  The
data were taken from Phase I (November 1999 - May 2001), when the
detector was filled with heavy water, and Phase II (July 2001 - August
2003), when NaCl was added to the target.
The search was a blind analysis in which the potential backgrounds
were estimated and analysis cuts were developed to eliminate such
backgrounds with $90\%$ confidence before the data were examined.  The
search maintained a greater than $50\%$ detection probability for
standard supernovae occuring at a distance of up to 60~kpc for Phase~I
and up to 70~kpc for Phase~II.  No low-multiplicity bursts were
observed during the data-taking period.
\end{abstract}


\section{Introduction}
Supernova neutrinos offer unique insights into both the fundamental nature of
neutrinos and the complex process of core collapse.  Although theoretical
calculations predict that a typical supernova releases approximately $3 \times
10^{53}$ ergs of gravitational binding energy, $99\%$ of which is carried away
by neutrinos, the only supernova neutrinos ever detected came from a single
supernova, SN 1987A~\citep{sn1987a_SK, sn1987a_imb, sn1987a_baksan}.  Many open questions in supernova core-collapse models
could be resolved with additional supernova neutrino data, motivating large
neutrino detectors to search their datasets for multiple events clustered
closely in time, which could be considered candidate supernova neutrino events.

One such neutrino detector, the Sudbury Neutrino Observatory (SNO)~\citep{snoDetector}, was an imaging water
Cherenkov detector located at a depth of 5890~m of water equivalent in the
Vale Inco., Ltd. Creighton mine near Sudbury, Ontario, Canada.  SNO detected neutrinos using
an ultra-pure heavy water ($^2$H$_2$O, hereafter D$_2$O) target contained in a
transparent spherical acrylic vessel 12~m in diameter.  Neutrino interactions in the vessel produced Cherenkov light that was detected by an array of 9456 photomultiplier tubes (PMTs) supported by a stainless steel geodesic structure.  Outside the geodesic structure were 5700 tonnes of light water that were monitored by outward-facing PMTs in order to identify cosmic-ray muons. 
The D$_2$O target
allowed SNO to detect neutrino events via three different reactions:
\begin{center}
  \begin{tabular}{ll}
     $ \nu_x + e^- \rightarrow \nu_x + e^-$  & (ES)\\
     $\nu_e + d \rightarrow p + p + e^-$\hspace{0.5in} & (CC)\\
     $ \nu_x + d \rightarrow p + n + \nu_x$ & (NC)\\  \\
  \end{tabular}
 \end{center}

	The neutral current (NC) reaction allowed SNO to detect all active neutrino flavors with equal sensitivity, while the charged current (CC)
reaction is exclusive to electron neutrinos.  The elastic scattering (ES) reaction is primarily sensitive to
electron neutrinos; other flavors can undergo ES reactions but with a smaller cross section.
Additionally, SNO could see electron antineutrinos through inverse $\beta$ decay on
deuterium and hydrogen, as well as through elastic scattering:
\begin{center}
  \begin{tabular}{ll}
     $\bar{\nu_e} + d \rightarrow n + n + e^+$\hspace{0.5in} & \\
     $\bar{\nu_e} + p \rightarrow  n + e^+$\hspace{0.5in} & \\
     $\bar{\nu_e} + e^{-} \rightarrow \bar{\nu_e} + e^{-}$\hspace{0.5in} & \\
  \end{tabular}
 \end{center}
The first reaction can provide a triple coincidence between the two neutrons
and the positron, but it has a relatively small cross section.  The second
reaction has a much higher cross section but occured only in SNO's 1700 tonne
light water shield, which was between the acrylic vessel and the PMT support structure. 
For this analysis we have focused only on the D$_2$O region.  Additional 
reactions on oxygen isotopes are also possible, but the rarity of $^{17}$O and
$^{18}$O in both the H$_2$O and D$_2$O volumes made the event rates from
these processes very low.  

	SNO's sensitivity to all neutrino flavors and the comparison of the
rates of the different possible reactions provide exciting opportunities to
distinguish various supernova models and investigate neutrino properties~\citep{flavor_sens_scholberg,flavor_sens_haxton,flavor_sens_schirato}.  During SNO's run time, it participated in the SuperNova Early Warning System
(SNEWS), which was designed to notify the astronomical community within minutes
of the detection of a large neutrino burst~\citep{snews}.  In order to avoid triggering on a false burst during this real time analysis, the threshold for the number of events qualifying as a burst was conservatively set to be 30 events in less than 2~s.  Although SNO did not observe such a large burst during its run time, a more thorough search is required to determine whether or not SNO observed any evidence of a supernova.

In this paper we have searched the SNO dataset for low-multiplicity bursts,
which we define as bursts of two or more events.  Such bursts could come from
supernovae in satellites of the Milky Way that are hidden by interstellar dust,
from non-standard supernovae in our own galaxy with relatively low neutrino
fluxes, or from completely unknown and unexpected sources of neutrinos.  The
Super-Kamiokande collaboration has recently published a similar
search~\citep{SuperK}.  Our search is complementary to that of Super-Kamiokande, in that for much of the running period examined here the Super-Kamiokande
detector was not operational.  Additionally SNO was primarily sensitive to electron neutrinos,
while Super-Kamiokande was primarily sensitive to electron antineutrinos.
%, and SNO also had the ability to differentiate between neutral current and charged current events.

\section{Data Analysis}

\subsection{Data Set}
The data analyzed here include two phases of SNO's operation.  Phase~I
ran from November 1999 to May 2001, and the sensitive volume of
the detector was filled only with D$_2$O.  Phase~II ran from July 2001 to
August 2003, and during this phase NaCl was added to the detector, increasing
the sensitivity to the NC reaction through the consequent enhancement of
neutron detection efficiency. Phase~II began
running shortly after Super-Kamiokande's first phase of data taking ended,
meaning that the majority of SNO Phase~II contains no overlap with the
supernova search performed by Super-Kamiokande.  Because of the enhanced NC
detection efficiency, SNO Phase~II provides a higher sensitivity to a
potential supernova signal.  The total livetime of Phase~I was 241.4 days, while the total livetime of Phase~II was 388.4 days.  The absolute time of an event was measured by a GPS system whose accuracy was $\sim 300$~ns~\citep{sno_phaseI}. 
%The time between events was provided by a GPS system with a resolution of 100~ns, and the absolute time of an event was known with an accuracy of $\sim 300$~ns~\citep{sno_phaseI}.  

\subsection{Search Parameters}
  We estimated all of the backgrounds that could mimic a supernova burst signal, and we designed our search windows and analysis cuts to ensure that we were $90\%$ confident we would not see a false burst.  We optimized our searches by adjusting the minimum burst multiplicity,
the length of the coincidence time interval, and the energy threshold for
individual events.  Our energy threshold sets a lower limit on the reconstructed total energy of the detected electrons, which may have significantly lower energies than the neutrinos themselves.  
Sensitivity to a supernova signal is increased by selecting a long coincidence time interval, a low multiplicity requirement, and a low energy threshold, 
%Higher sensitivity is obtained for the
%lowest-multiplicity, longest, lowest-threshold window, 
but these criteria must be balanced against the increase 
in the probability of a
``background'' burst, especially through accidental coincidences. 
As the energy threshold is lowered, the event rate goes up, and hence the
accidental coincidence rate increases, particularly for
low-multiplicity bursts and for a long coincidence time interval.  To ensure a
90$\%$ probability of seeing no false bursts within a given search window, the expected backgrounds from all sources must sum to no more than 0.11 events.
%our expected backgrounds
%from all sources must sum to an average of no more than 0.11 events for the livetime of the data phase.
%Accidentals are especially dangerous because their uniform position
%distribution within the detector can mimic a supernova. 
Our optimization ultimately led us to perform three distinct searches: two
multiplicity two ($N_{\rm event}=2$) searches, and one multiplicity
three ($N_{\rm event}=3$) search.

The first of our $N_{\rm event}=2$ searches used a short window
(0.05~s) to focus on detecting neutrinos from a supernova
neutronization burst.  In the case of a failed supernova, the
neutronization burst provides the only potential signal because
shortly after the neutronization phase, the supernova collapses into a
black hole, abruptly terminating the neutrino signal~\citep{beacom_failedSN, failedSN_mclaughlin,failedSN}.  These unusual supernovae are of special interest to
astronomers, and their neutrino signatures could provide interesting
model constraints.  The second multiplicity two window was of moderate
length (0.2~s for Phase~I and 1~s for Phase~II) to maximize our
sensitivity to a standard supernova event.  Table~\ref{tbl:windows}
summarizes the search windows and their respective energy thresholds.

\begin{deluxetable}{ccc}
\tablewidth{0pt}
\tablecaption{Search windows and energy thresholds}
\tablenum{1}
\tablehead{\colhead{} & \colhead{Window Length} & \colhead{Energy Threshold} \\ 
\colhead{} & \colhead{(s)} & \colhead{(MeV)} } 

\startdata
Phase~I, N$_{\rm{event}}=2$ & 0.05 & 5.0 \\
Phase~I, N$_{\rm{event}}=2$ & 0.2 & 6.0 \\
Phase~I, N$_{\rm{event}}=3$ & 10.0 & 4.5 \\
Phase~II, N$_{\rm{event}}=2$ & 0.05 & 6.5 \\
Phase~II, N$_{\rm{event}}=2$ & 1.0 & 8.5 \\
Phase~II, N$_{\rm{event}}=3$ & 10.0 & 4.5 \\
\enddata
\tablecomments{The above values were chosen to maximize our sensitivity while limiting our probability of seeing a false burst.  The energy threshold sets a lower limit on the reconstructed total energy of the detected electrons.\label{tbl:windows}}
\end{deluxetable}

The Phase~I data set entirely overlaps with the running of
Super-Kamiokande, while much of Phase~II does not overlap.
Consequently for Phase~I our optimization was intended to maintain
some neutral current sensitivity, which Super-Kamiokande does not
have to great extent, with the aim of being able to detect non-standard burst
sources. For Phase~II, however, our goal was to maximize overall
supernova sensitivity, and therefore we used a fairly large (1~s)
search window.  The 1~s window required us to raise the energy
threshold high enough that there is very little remaining neutral
current signal, but it increased our overall sensitivity to standard
supernovae bursts.

For the multiplicity three search, where accidentals are not a major
background, the window for each phase is 10~s, and the energy threshold is 4.5
MeV.  Primarily because of the low energy threshold available for this search,
it provides the best sensitivity to standard supernovae bursts.  

In order to simulate a ``standard'' supernova burst, we utilized supernova Monte Carlo simulations based on the Burrows model~\citep{BurrowsModel}.  The average neutrino energies from this model are $\langle E_{\nu_e} \rangle \simeq 13$~MeV, $\langle E_{\overline{\nu}_e} \rangle \simeq 15.5$~MeV, and $\langle E_{''\nu_\mu''} \rangle \simeq 20$~MeV, where $''\nu_\mu''$ represents any of the flavors $\nu_\mu$, $\overline{\nu}_\mu$, $\nu_\tau$, or $\overline{\nu}_\tau$~\citep{BurrowsDetail}.

\subsection{Backgrounds}

The primary difficulty in performing a triggerless burst search is the
elimination of the ``background'' created by false bursts.  In addition to the accidental
coincidences, a large number of correlated physics backgrounds
must be estimated and almost entirely eliminated.  SNO's sensitivity to neutrons makes this
problem particularly difficult: any process that produces multiple neutrons
can lead to an apparent burst, with the average time between neutron captures being
roughly 50~ms in Phase~I due to capture on deuterium and 5~ms in Phase~II due to capture on Cl.  Table~\ref{tbl:bkgd} shows our background
estimates for each of our search windows before we apply any special analysis cuts designed to remove false bursts beyond the standard SNO analysis cuts~\citep{leta, sno_phaseI}.  Most of the multiplicity three
backgrounds are conservatively assumed to have upper limits corresponding to
the multiplicity two estimates.  
%The limits were added assuming they were
%90$\%$ confidence interval fluctuations of Poissonian processes.

The dominant correlated backgrounds are interactions by  atmospheric neutrinos,
which can produce neutrons without any primary large energy deposit to tag the
events.  For both Phase~I and Phase~II, the atmospheric background was
estimated with the neutrino interaction generator NUANCE~\citep{nuance}, whose
output was then further processed by the full SNO detector simulation.  The simulation's atmospheric neutrino energies ranged from 100~MeV to 2~TeV, and flavor oscillation corrections were applied.  The systematic error in the NUANCE simulation is conservatively estimated to be $\pm$20$\%$, and it is dominated by uncertainties in the neutrino cross sections. %Table~\ref{tbl:bkgd} shows how widely the atmospheric background estimations vary for different search windows and energy thresholds.

Most muons traveling through the SNO detector were tagged by outward-looking
PMTs, and neutrons following these muons are eliminated by a 20~s software
`muon follower' veto.  Some muons, however, do not have enough energy to
trigger the outward-looking PMTs and will leak into the detector.  
Fortunately few of these muons are likely to produce multiple neutrons that could mimic a burst.  In the entire livetime of Phase~II, the effect of the remaining leaked muons
is estimated to cause fewer than 1.35 single-neutron events, implying a
conservative upper limit of fewer than 0.5 coincidence events for Phase~II.  Since Phase~I would have even fewer coincidences, the same conservative upper limit of 0.5 coincidences is assumed.

Although the SNO detector was constructed from materials with low radioactivity, some radioactive
backgrounds still existed, such as $^{238}$U, and spontaneous fission from
residual radioactivity can lead to false bursts due to multiple neutron
capture.  Many of the radioactive backgrounds discussed in previous SNO analyses are not significant in this search because they will not produce bursts of events, but fission neutrons from $^{238}$U can create a background burst.  
The amount of $^{238}$U in the detector was measured using an inductively coupled plasma mass spectrometer in September
2003 after the addition of NaCl to the heavy water~\citep{sno_nuebar_search}.  
%This measurement set an
%upper limit of 0.79 coincidences for Phase~I due to spontaneous fission from
%$^{238}$U.  In Phase~II, the addition of NaCl prevented the use of 
%SNO's reverse osmosis purification system, meaning that more $^{238}$U could
%have been present.  For this phase a very conservative upper limit of 10 fission bursts due to $^{238}$U was estimated based upon the standard probability of the fission producing various neutron multiplicities~\citep{fission}.  
More $^{238}$U could have been present in Phase~II than in Phase~I because the addition of NaCl prevented the use of SNO's reverse osmosis purification system; therefore, this $^{238}$U measurement was used as an upper limit for Phase~I.  Using the standard probability of a fission producing various neutron multiplicities~\citep{fission}, we estimated a very conservative upper limit of 0.79 (10) fission bursts due to $^{238}$U for Phase~I (Phase~II).
This limit also assumed a 65$\%$ detection efficiency for neutrons above 4.5~MeV and a 40$\%$ efficiency for detecting a gamma burst in coincidence with the fission.

	A very small fraction of the time, the production of even a single
neutron via the NC reaction can lead to an apparent burst.  If a $\gamma$ ray
from the capture on deuterium (in Phase~I) or chlorine (Phase~II) produces an
electron above threshold via Compton scattering and then subsequently photodisintegrates a
deuteron, it can create a second neutron.  Based on the SNO Monte Carlo simulation,
this photodisintegration background has been estimated to cause a small background
of less than 0.01 coincidences in Phase~I and $0.43 \pm 0.03$ coincidences in Phase~II.

	Antineutrinos can also lead to apparent bursts, due to the primary
positron created in the interaction and the following capture of one or two
neutrons.  Antineutrinos from radioactive nuclei in the earth surrounding
the detector lead to a negligible background from this process for Phase~I and $0.5 \pm 0.1$ bursts for Phase~II.  A study of all
commercial nuclear reactors within 500~km of SNO determined that the
coincidence background from antineutrinos from these reactors was also very
small, $0.019 \pm 0.002$ coincidences for Phase~I and $1.4 \pm 0.3$ for Phase~II.  

	During the SNO detector construction the acrylic was exposed to air
containing radon.  In the decay chain of radon is $^{210}$Po, which can decay via $\alpha$-emission.  The $\alpha$ can interact with the carbon in the acrylic, leading to $^{13}$C$(\alpha, n)^{16}$O reactions in which the $^{16}$O will produce a $e^{+}e^{-}$ pair or a $\gamma$-ray that can photodisintegrate deuterium.  The estimates of $(\alpha,n)$ 
coincidences from Monte Carlo studies of this background are low.  Coincidences due to the diffuse supernova background (DSNB) and to instrumental
background events are also estimated to be quite low, as shown in
Table~\ref{tbl:bkgd}.

\begin{deluxetable}{lcccccc}
\tablewidth{0pt}
\tabletypesize{\scriptsize}
%\tablewidth{0.9\textwidth}
\tablecaption{Physics backgrounds that could create false bursts} 
\tablenum{2}
\tablehead{\colhead{Backgrounds} & \colhead{Phase~I} & \colhead{Phase~II} & \colhead{Phase~I} & \colhead{Phase~II} & \colhead{Phase~I} & \colhead{Phase~II} \\ 
\colhead{} & \colhead{N$_{\rm event}=2$} & \colhead{N$_{\rm event}=2$} & \colhead{N$_{\rm event}=2$} & \colhead{N$_{\rm event}=2$} & \colhead{N$_{\rm event}=3$} & \colhead{N$_{\rm event}=3$} \\
\colhead{} & \colhead{0.05s window} & \colhead{0.05s window} & \colhead{0.2s window} & \colhead{1s window} & \colhead{10s window} & \colhead{10s window} }

\startdata
Atmospherics & 4.5 $\pm$ 0.9 & 6.9 $\pm$ 1.4 & 1.9 $\pm$ 0.4 & 0.6 $\pm$ 0.1 & 1.5 $\pm$ 0.3 & 7.5 $\pm$ 1.5\\ 
Muon Spallation & $< 0.5$ & $< 0.5$ & $< 0.5$ & $< 0.5$ & $< 0.5$ & $<0.5$\\ 
Fission $^{238}$U & $< 0.8$ & $<10$ & $< 0.8$ & $< 10$ & $< 0.8$ & $< 3$ \\ 
Photodisintegration & $< 0.01$ & $0.4 \pm 0.03$ & $< 0.01$ & $0.4 \pm 0.03$ & $< 0.01$ & $< 0.4$\\ 
geo-$\nu$ & 0.0 & $0.5 \pm 0.1$ & 0.0 & $0.5 \pm 0.1$ & 0.0 & $< 0.5$\\ 
reactor-$\nu$ & 0.02 $\pm 0.002$ & $1.4 \pm 0.3$ &0.02 $\pm 0.002$ & $1.4 \pm 0.3$ & $< 0.02$ & $< 1.4$\\ 
($\alpha$,$ne^{+}e^{-}$) & 0.02 $\pm 0.10$ & 0.07 $\pm 0.07$ &0.02 $\pm 0.10$ & 0.07 $\pm 0.07$ & $< 0.02$ & $< 0.07$ \\ 
($n$,$2n$) & $<0.02$ & $< 0.07$ & $< 0.02$ & $< 0.07$ & $< 0.02$ & $< 0.07$ \\ 
DSNB & $< 0.005$ & $< 0.005$ & $< 0.005$ & $< 0.005$ & $< 0.005$ & $< 0.005$\\ 
Instrumentals & $< 0.03$ & $< 1$ & $< 0.03$ & $< 1$ & $< 0.03$ &$< 1$\\ 
%\hline
%Total Backgrounds & $5.5 \pm 1.3$ & $15.7 \pm 3.3$ & $ 2.9 \pm 1.1$ & $9.3 \pm 3.0$ & $1.7 \pm 0.6$ & $9.5 \pm 2.1$ \\
\enddata
\tablecomments{Since the search windows have different energy thresholds, we estimate the backgrounds for each search window for both Phase~I and Phase~II.  These backgrounds are further reduced by the analysis cuts described in the text.\label{tbl:bkgd}} 
\end{deluxetable}

\subsection{Analysis Cuts}

	We developed a set of analysis cuts, beyond the standard cuts used by
other SNO analyses~\citep{leta, sno_phaseI}, to reduce the level of correlated backgrounds
shown in Table~\ref{tbl:bkgd} and discussed in the previous section. As in previous analyses, we utilized a fiducial volume radius of 550~cm, as well as a variety of instrumental cuts based on PMT charge and timing information.  Our high level cuts incorporated information such as the isotropy of the detected light and the event's reconstruction quality, but we did not include any of the cuts SNO previously designed to remove bursts from the data set.  Instead we designed new cuts that could discriminate between background bursts and potential supernova bursts.

	False bursts from atmospheric reactions often have a high-energy
primary, followed by delayed neutron captures.  We removed these bursts by
imposing a deadtime window following any event whose energy exceeded roughly
80~MeV.  The 80~MeV threshold was chosen to minimize the acceptance loss
for neutrinos from a standard supernova, whose energies tend to peak near
20~MeV.  For Phase~I, in which the neutron capture time is $\sim 50$~ms, we
used a deadtime window of $600$~ms, and for Phase~II, in which the capture time
is $\sim5$~ms, we used a deadtime window of 200~ms.  When we applied this cut to our Monte Carlo simulation of a standard supernova, no genuine supernova bursts were removed.
%SNO generated photon detection probabilities based on the reconstructed event position and direction, and it utilized all the detected PMT hits in making its kinetic energy estimate.  The best value of the kinetic energy was found by maximizing the likelihood given the observed number of hit PMTs, taking into account optical effects from the reconstructed position and direction of the event~\citep{leta}.

	In addition to having large energies, the primaries from an atmospheric
neutrino interaction often have multiple tracks or are heavier particles than
the electrons expected either from charged current supernova interactions or
from the Compton scattering of $\gamma$ rays released in neutron capture.  We
removed false bursts associated with these events by tagging any event that was
not electron-like, and removing all events within 600~ms (Phase~I) and 200~ms
(Phase~II) afterward.  Our definition of an electron-like event was based on 
%We defined electron-like  based on the isotropy of the
the PMT hit pattern and Kolgmogorov-Smirnov tests on the angular
distribution of Cherenkov light of reconstructed events, as has been done in
other SNO analyses~\citep{leta}.  We estimated the acceptance of this cut by
applying it to our Monte Carlo simulation of supernova bursts, and we found that
for both phases of the experiment, only $1.2\%$ of genuine supernova bursts
were removed.  The same cut applied to simulated atmospheric neutrino events
removed $57\%$ of those bursts that pass the standard analysis cuts from Phase~I and $63\%$ from Phase~II.

\begin{figure}[H]
\begin{center}
%{\rotatebox{0}{\resizebox{3.1in}{!}{\includegraphics{plots/drdt_4in1_paper_plot_small.eps}}}}
{\rotatebox{0}{\resizebox{0.45\textwidth}{!}{\includegraphics{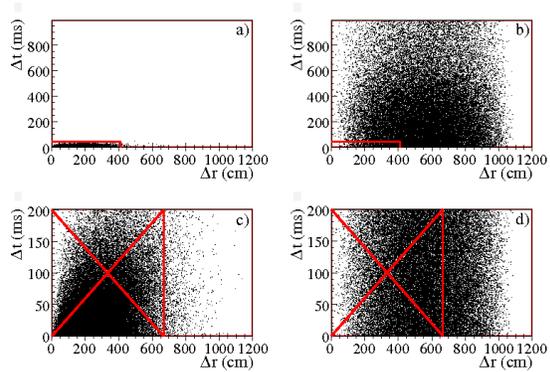}}}}
\caption{a) Simulated background bursts in the 1~s search window for Phase II: The region inside the box was removed by the $\Delta$r$\Delta$t cut, which eliminates most of the background events; b) Simulated supernova signal in the 1~s window for Phase II: The $\Delta$r$\Delta$t cut removes very little of the simulated signal; c) Simulated background bursts in the 0.2~s search window for Phase I:  The region marked with an X was removed by the $\Delta$r cut, which eliminates most of the background events; d) Simulated supernova signal in the 0.2~s search window for Phase I: The $\Delta$r cut must remove a large fraction of the supernova signal in order to sufficiently eliminate background bursts.\label{fig:drdt}}
%\caption{a) The region inside the box was removed by the $\Delta$r$\Delta$t cut, which eliminates most of the simulated background bursts shown here for the Phase~II 1~s search window b) The $\Delta$r$\Delta$t cut removed very little of the simulated supernova signal shown here for the Phase~II 1~s window c) The region marked with an X was removed by the $\Delta$r cut, which eliminates most of the simulated background bursts shown here for the Phase~I 0.2~s window d) The $\Delta$r cut in Phase~I must remove a large fraction of the simulated supernova signal in order to sufficiently eliminate background bursts.\label{fig:drdt}}
%\caption{ a) Example $\Delta$r$\Delta$t cut applied to simulations of our expected background bursts for the Phase~II 1~s window search.  b) Example $\Delta$r$\Delta$t cut applied to simulations of the expected signal from a standard supernova seen in the Phase~II 1~s window. c) The X shows an example region removed by the $\Delta$r cut applied to simulations of our expected background bursts for the Phase~I 0.2~s window search d) Example $\Delta$r cut applied to simulations of the expected signal from a standard supernova seen in the Phase~I 0.2~s window.  \label{fig:drdt}}
\end{center}
\end{figure}

	With the exception of accidental coincidences, almost all sources of false
bursts are spatially correlated.  Neutrons from atmospheric interactions or
fission will capture near the primary event and near one another.  To remove events
on this basis, we use a cut developed by the Super-Kamiokande collaboration in
their triggerless burst analysis~\citep{SuperK}. 
We define $\Delta$r as a weighted mean of the distances between the reconstructed event
positions  in a candidate burst:
\begin{equation}\label{eq_deltaR}
{\Delta}r = \frac{{\Sigma}^{M-1}_{i=1}{\Sigma}^{M}_{j=i+1}|\vec{r_{i}}
- \vec{r_{j}}|}{_{M}C_{2}}
\end{equation}
where $|\vec{r_{i}} - \vec{r_{j}}|$ is the distance between the
reconstructed positions of events $i$ and $j$ within a burst, $M$ is
the multiplicity of the burst, and $_{M}C_{2}$ is the number of
non-redundant combinations.  For SNO, we can improve on the straightforward 
$\Delta$r cut by using the fact that most spatially correlated events are due
to neutron captures, and therefore there is a direct relationship between $\Delta$r of the events and the time that separates them, due to the
neutron diffusion time within the heavy water.  We therefore examined not just
the spatial separation of burst events but also their separation in time,
$\Delta$t, which is defined similarly to Eq.~\ref{eq_deltaR}.  Figure~\ref{fig:drdt} shows a $\Delta$r$\Delta$t distribution for $N_{\rm
event}=2$ from Phase~II
where the two-dimensional cut is clearly beneficial.  The figure also shows the
same distribution for Phase~I, where the longer neutron capture time and longer
diffusion distance makes the two-dimensional cut less effective and hence we
use only the cut on $\Delta$r.  For both phases
we applied the two-dimensional cut to the $N_{\rm event}=3$ search.  Table~\ref{tbl:drdt} summarizes the fraction of expected supernova signal that survives the $\Delta$r or $\Delta$r$\Delta$t cut for each of the search windows.  
%The functional form of the $\Delta$r$\Delta$t cut was generally box-shaped as described in Table~\ref{tbl:drdt}, though for most of the search windows the shape of the cut included a rounded corner.  
The functional form of the $\Delta$r$\Delta$t cut was generally box-shaped, though for several of the search windows the shape of the cut included a rounded corner described by $\Delta$t $> \sqrt{\alpha_{1}^2 + \alpha_{2}\alpha_{3} -\alpha_{2}\Delta{\mathrm{r}}}$.  Table~\ref{tbl:drdt} shows the parameter values used for the $\Delta$r$\Delta$t cut for each of the search windows.
In Phase~I the long diffusion distance makes eliminating backgrounds more difficult, meaning that the $\Delta$r cut must be harsher in order to adequately reduce the probability of seeing a false burst.

%Table~\ref{tbl:drdt} also indicates the placement of the $\Delta$r part of the cut, though the functional form of the $\Delta$r$\Delta$t cut varies for different search windows, meaning that the $\Delta$r$\Delta$t cut is not always defined by straight lines.   

\begin{deluxetable}{lcccccc}
%\tablewidth{0.45\textwidth}
\tablewidth{0pt}
\tablecaption{Fraction of standard SN signal that survives the $\Delta$r$\Delta$t cut}
\tablenum{3}
\tablehead{\colhead{} & \colhead{$\Delta$r Cut} & \colhead{$\Delta$t Cut} & \colhead{$\alpha_1$} & \colhead{$\alpha_2$} & \colhead{$\alpha_3$} & \colhead{SN Signal} \\ 
\colhead{} & \colhead{(cm)} & \colhead{(ms)} & \colhead{(ms)} & \colhead{(ms)$^2$/(cm)} &\colhead{(cm)} & \colhead{} }  

\startdata
Phase~I, 0.05s window & 626 & -- & -- & -- & -- & 0.389 \\
Phase~I, 0.2s window & 668 & -- & -- & -- & -- & 0.303 \\
Phase~I, 10s window & 600 & 180 & 180 & 100 & 376 & 0.867 \\
Phase~II, 0.05s window & 385 & 40 & 40 & 5 & 200 & 0.811 \\
Phase~II, 1s window & 410 & 45 & -- & -- & -- & 0.977 \\
Phase~II, 10s window & 400 & 35 & 35 & 10 & 300 & 0.999 \\
\enddata
\tablecomments{Only the $\Delta$r cut was applied to the N$_{\rm event}=2$ searches in Phase~I, while we applied the $\Delta$r$\Delta$t cut to Phase~II and to the N$_{\rm event}=3$ search in Phase~I. The $\Delta$r$\Delta$t cuts were generally box-shaped with one rounded corner described by the $\alpha$ parameters.  \label{tbl:drdt}}
\end{deluxetable}

\section{Results}

	After estimating all of the background sources of false bursts that could mimic a supernova burst signal, we placed our analysis cuts to ensure that we expected no more than 0.11 events in each search window.  Our approach to the low-multiplicity search was inherently blind: we did not examine the real data
until we had set all of our search parameters, and we did not change any of
those parameters after we performed the search.  

Our $\Delta$r$\Delta$t cut defined an ``antibox'' region that was excluded from the supernova burst search, as shown in Fig~\ref{fig:drdt}.
Prior to opening the box and examining the events passing the analysis cuts, we examined the antibox in order to confirm our estimates of background bursts.  We found approximately the same number of bursts there
that we expected, as shown in Table~\ref{tbl:antibox_results}.  Because our background was dominated by atmospheric neutrino bursts and several of our other background estimates were conservative upper limits, Table~\ref{tbl:antibox_results} compares the number of bursts we observed outside the box to the number of bursts we expected due to atmospheric events.

	After all the cuts were developed and tested on simulations, we fixed our
analysis and performed our burst searches on both the Phase~I and Phase~II
data. We observed no bursts in any of our search windows, as summarized in Table~\ref{tbl:box_results}.

\begin{figure}[H]
\begin{center}
%{\rotatebox{0}{\resizebox{3.1in}{!}{\includegraphics{plots/box_deltaRdeltaT_sn_salt_10s_combo.eps}}}}
{\rotatebox{0}{\resizebox{0.45\textwidth}{!}{\includegraphics{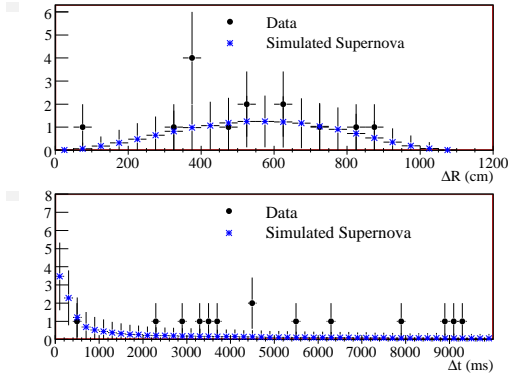}}}}
\caption{$\Delta$r and $\Delta$t distributions for $N_{\rm event}=2$ bursts found in the $N_{\rm event}=3$ search window for Phase~II.  These bursts are consistent with expected accidental coincidences and are not considered candidate supernova bursts.\label{fig:drdtn2}}
\end{center}
\end{figure}

Because our 10~s window search was optimized for an $N_{\rm event}=3$ burst, we did observe some $N_{\rm event}=2$ bursts in that window.  Using the same energy threshold and analysis cuts designed for the Phase~II 10~s $N_{\rm event}=3$ search, we observed 14 $N_{\rm event}=2$ bursts, which is in keeping with our expectations from
accidental coincidences.  Figure~\ref{fig:drdtn2} shows the distribution of
$\Delta$r and $\Delta$t for these $N_{\rm event}=2$ bursts.  The distribution
in $\Delta$r is approximately uniform,
which is consistent with the hypothesis that these are accidental bursts, though it is also consistent with the hypothesis that these are supernova bursts.  The $\Delta$t distribution supports the conclusion that these are accidental bursts since the events are spread evently throughout the time window, in contrast with our expectations from an actual supernova.
%which is consistent with the hypothesis that these are accidental bursts.  
%The $\Delta$t distribution further supports this conclusion since the events are spread evenly throughout the time window, in contrast with our expectations from an actual supernova. 
The event with
the lowest $\Delta$t separation also has a $\Delta$r too low to have survived
any of the $\Delta$r$\Delta$t cuts for the multiplicity two searches, meaning that it is more likely to have been a background burst than a genuine supernova burst. We also
observed two $N_{\rm event}=2$ bursts in the $N_{\rm event}=3$ search for 
Phase~I.  These two bursts are separated by 3.2~s and 8.0~s respectively,
which puts them well outside of the $N_{\rm event}=2$  search windows.  These
bursts are also consistent with our accidental coincidence expectations. 

\begin{deluxetable}{lcc}\label{tbl:antibox_results}
\tablewidth{0pt}
\tablecaption{Low-multiplicity bursts in the background-contaminated region}
\tablenum{4}
\tablehead{\colhead{Search description} & \colhead{Bursts expected } & \colhead{Bursts found} \\ 
\colhead{} & \colhead{in $\Delta$r$\Delta$t} & \colhead{in $\Delta$r$\Delta$t} \\
\colhead{} & \colhead{excluded region} & \colhead{excluded region} 
 } 

\startdata
Phase~I, 0.05s window & $4.53 \pm 0.91$& 2 \\
Phase~I, 0.2s window & $1.90 \pm 0.38$& 1 \\
Phase~I, 10s window & $1.51 \pm 0.30$& 2 \\
Phase~II, 0.05s window & $6.94 \pm 1.39$& 7 \\
Phase~II, 1s window & $0.57 \pm 0.11$& 1 \\
Phase~II, 10s window & $7.51 \pm 1.50$& 9 \\
\enddata
%\tablecomments{In the $\Delta$r$\Delta$t regions that we excluded from our search due to heavy background contamination, we found burst levels that were consistent with our background estimates, which are dominated by the atmospheric neutrino burst estimates shown in this table.\label{tbl:antibox_results}}
\end{deluxetable}

\begin{deluxetable}{lcc}
\tablewidth{0pt}
\tablecaption{Low-multiplicity bursts found in the SNO dataset}
\tablenum{5}
\tablehead{\colhead{Search description} & \colhead{Bursts expected } & \colhead{Bursts found} \\ 
\colhead{} & \colhead{in search region} & \colhead{in search region} \\ 
 }

\startdata
Phase~I, 0.05s window & $<0.11$ & 0 \\
Phase~I, 0.2s window & $< 0.11$ & 0 \\
Phase~I, 10s window & $<0.11$ & 0 \\
Phase~II, 0.05s window & $<0.11$ & 0 \\
Phase~II, 1s window & $<0.11$ & 0 \\
Phase~II, 10s window & $<0.05$ & 0 \\
\enddata
\tablecomments{We observed no candidate bursts in any of the search windows.\label{tbl:box_results}}
\end{deluxetable}

\section{Discussion and Conclusions}

We performed a triggerless search for low-multiplicity bursts in 
data from Phase~I (D$_2$O only) and Phase
II (D$_2$O loaded with NaCl) of the Sudbury Neutrino Observatory, finding no
candidate bursts.  For the period of overlap our results are consistent with the null signal observed by
Super-Kamiokande. 

\begin{figure}[H]
\begin{center}
%{\rotatebox{0}{\resizebox{3.1in}{!}{\includegraphics{plots/sn_sens_d2o.eps}}}}
{\rotatebox{0}{\resizebox{0.45\textwidth}{!}{\includegraphics{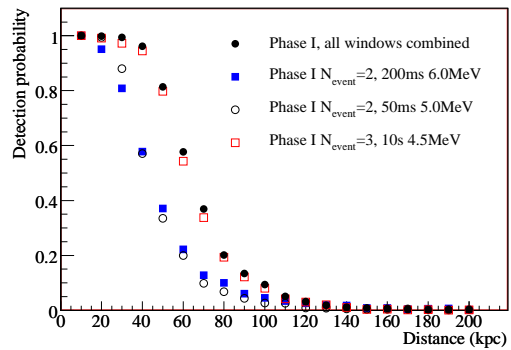}}}}
\caption{Probability of detecting a supernova in Phase~I data assuming a standard Burrows model.\label{fig:sens_d2o}}
\end{center}
\end{figure}

\begin{figure}[H]
\begin{center}
%{\rotatebox{0}{\resizebox{3.1in}{!}{\includegraphics{plots/sn_sens_salt.eps}}}}
{\rotatebox{0}{\resizebox{0.45\textwidth}{!}{\includegraphics{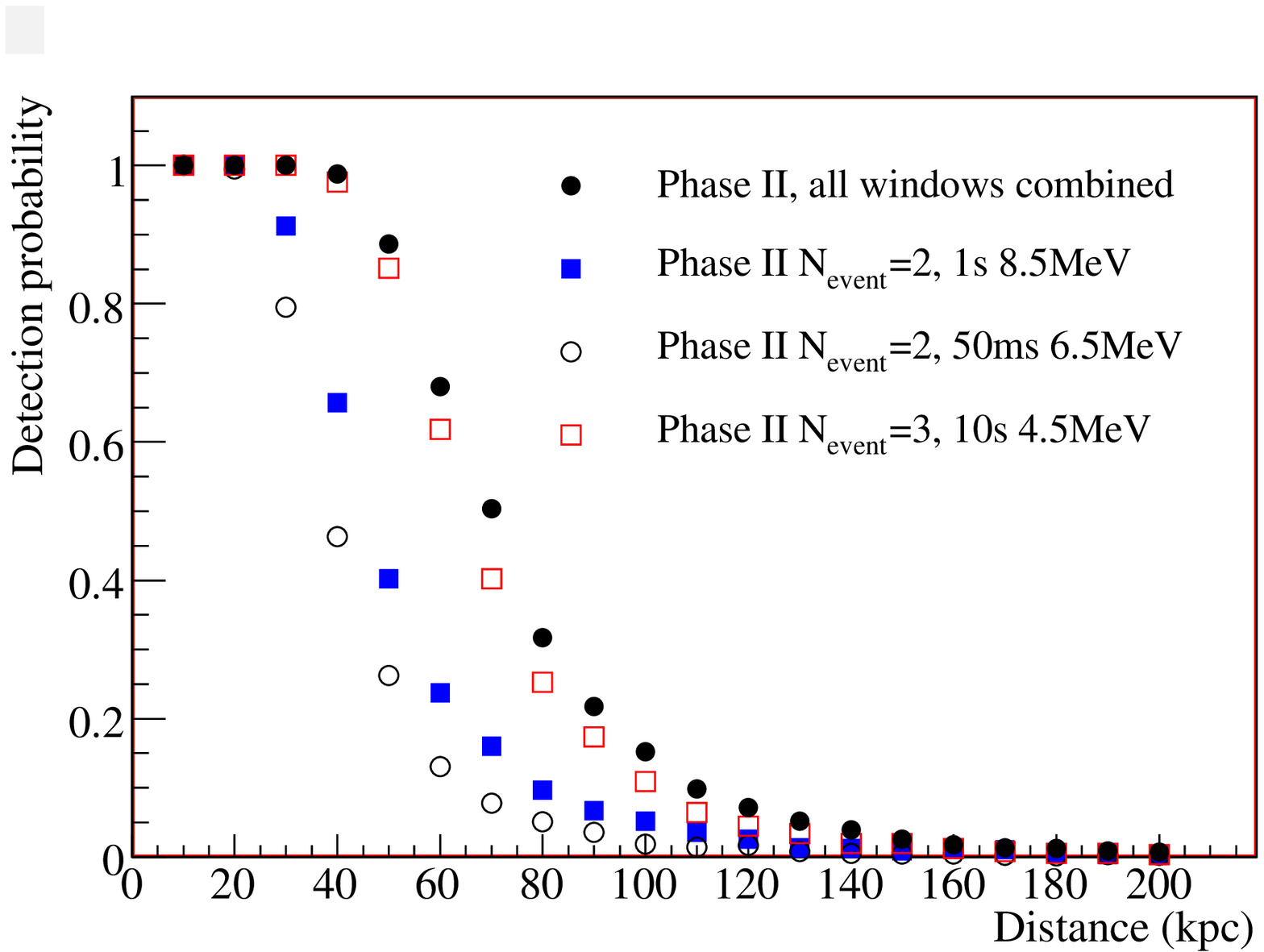}}}}
\caption{Probability of detecting a supernova in Phase~II data assuming a standard Burrows model.\label{fig:sens_salt}}
\end{center}
\end{figure}

Figures~\ref{fig:sens_d2o}-\ref{fig:sens_salt} show the sensitivity of our various search windows to a standard supernova for
both phases as a function of supernova distance.  Those figures do not include the small dispersion effects arising from the nonzero mass of the neutrino.
%In order to estimate the
%probability of supernova detection, we utilized supernova Monte Carlo simulations based on
%the standard Burrows model~\citep{BurrowsModel}.  
For Phase~I, which
was completely overlapped by the run time of Super-Kamiokande, our search was
primarily looking for non-standard supernova signals in which $\nu_e$ emission
would be suppressed, allowing SNO to detect a neutral current signal that Super
Kamiokande might not have seen.  

At a typical distance in our galaxy, 10~kpc, we retain a $100\%$ detection probablity for a standard core-collapse supernova. 
%as well as for a failed supernova that emits only a neutronization burst.  
In Phase~I we maintain a $50\%$ detection probability for a standard supernova out to 60~kpc.  In Phase~II we retain a 100$\%$ detection probability out to 30~kpc and a greater than $50\%$ detection probability out to 70~kpc.  

\section{Acknowledgments}
 This research was supported by: Canada: Natural Sciences and
Engineering Research Council, Industry Canada, National Research
Council, Northern Ontario Heritage Fund, Atomic Energy of Canada,
Ltd., Ontario Power Generation, High Performance Computing Virtual
Laboratory, Canada Foundation for Innovation, Canada Research Chairs;
US: Department of Energy, National Energy Research Scientific
Computing Center, Alfred P. Sloan Foundation; UK: Science and
Technology Facilities Council; Portugal: Funda\c{c}\={a}o para a
Ci\^{e}ncia e a Technologia.  We thank the SNO technical staff for
their strong contributions.

\bibliographystyle{plainnat}

\end{document}